\documentclass[conference,compsoc]{IEEEtran}
\IEEEoverridecommandlockouts

\usepackage[T1]{fontenc}

\AtBeginDocument{%
  \providecommand\BibTeX{{%
    \normalfont B\kern-0.5em{\scshape i\kern-0.25em b}\kern-0.8em\TeX}}}

\usepackage{xurl}
\usepackage{nicefrac}
\usepackage{siunitx}
\usepackage{array,framed}
\usepackage{booktabs}
\usepackage{paralist}
\usepackage{
  color,
  float,
  epsfig,
  wrapfig,
  graphics,
  graphicx,
  subcaption
}
\usepackage{textcomp,amssymb,natbib}
\usepackage{setspace}
\usepackage{latexsym,fancyhdr,url}
\usepackage{enumerate}
\usepackage{algorithm2e}
\usepackage{algpseudocode}
\usepackage{graphics}
\usepackage{xparse}
\usepackage{xspace}
\usepackage{multirow}
\usepackage{csvsimple}
\usepackage{balance}
\usepackage{tabularx}
\usepackage{enumitem}
\usepackage{caption}
\usepackage{placeins}
\usepackage{hyperref}
\usepackage{mathtools}
\usepackage{cleveref}
\usepackage{etoolbox}
\newcommand{\startappendices}{%
  \appendices
  \crefalias{section}{appendix}%
  \crefname{appendix}{Appendix}{Appendices}%
  \Crefname{appendix}{Appendix}{Appendices}%
}

\crefformat{figure}{#2Figure~#1#3}
\Crefformat{figure}{#2Figure~#1#3}
\crefformat{table}{#2Table~#1#3}
\Crefformat{table}{#2Table~#1#3}
\crefformat{appendix}{#2Appendix~#1#3}
\Crefformat{appendix}{#2Appendix~#1#3}

\newcommand{\BfPara}[1]{\vspace{1mm}{\noindent\bf#1.}\xspace\xspace}

\usepackage{pifont}
\newcommand{\circleone}{%
\footnotesize
  \begin{tikzpicture}[baseline=-0.4em]
        \node[left,circle,minimum size=10,inner sep=0,outer sep=3,draw] {1};
  \end{tikzpicture}\normalsize}
\newcommand{\circletwo}{%
\footnotesize
  \begin{tikzpicture}[baseline=-0.4em]
        \node[left,circle,minimum size=10,inner sep=0,outer sep=3,draw] {2};
  \end{tikzpicture}\normalsize}
\newcommand{\circlethree}{%
\footnotesize
  \begin{tikzpicture}[baseline=-0.4em]
        \node[left,circle,minimum size=10,inner sep=0,outer sep=3,draw] {3};
  \end{tikzpicture}\normalsize}
\newcommand{\circlefour}{%
\footnotesize
  \begin{tikzpicture}[baseline=-0.4em]
        \node[left,circle,minimum size=10,inner sep=0,outer sep=3,draw] {4};
  \end{tikzpicture}\normalsize}

\hypersetup{
    pdfpagemode=pagewidth,
    plainpages=false,
    colorlinks,
    urlcolor=blue!70!black,
    linkcolor=red!70!black,
    citecolor=green!70!black,
}

\usepackage{quoting}
\quotingsetup{
  vskip=0.5em,
  leftmargin=1em,
  rightmargin=1em
}
  
\usepackage[framemethod=TikZ]{mdframed}
\definecolor{rubcolor}{HTML}{7E8995}
\global\mdfdefinestyle{insightstyle}{%
backgroundcolor=gray!10,
outerlinewidth=1pt,innerlinewidth=0pt,
outerlinecolor=black,roundcorner=5pt
}
\newmdenv[roundcorner=10pt, frametitle=Key Takeaways, linecolor=rubcolor]{insightbox}

\usepackage{
  tikz,
  pgfplots,
  pgfplotstable
}

\usepackage{svg}

\usetikzlibrary{
  shapes.geometric,
  arrows,
  external,
  pgfplots.groupplots,
  matrix,
  positioning,
  automata,
  fit
}

\pgfplotsset{compat=1.9}

\usepackage{mathtools,}

\DeclareMathAlphabet{\mathcal}{OMS}{cmsy}{m}{n}
\DeclareGraphicsExtensions{%
    .png,.PNG,%
    .pdf,.PDF,%
    .jpg,.mps,.jpeg,.jbig2,.jb2,.JPG,.JPEG,.JBIG2,.JB2}

\begin{document}

\title{Achievement Unlocked: Let’s Get Hacked! \\ An Empirical Study of Cybercrime in the Video Gaming Ecosystem}

\author{
\IEEEauthorblockN{
Janine Schneider\IEEEauthorrefmark{1},
Jan Kallenborn \IEEEauthorrefmark{2},
Tim Hoffmann\IEEEauthorrefmark{3},\\
Maximilian Eichhorn\IEEEauthorrefmark{3},
Thorsten Holz\IEEEauthorrefmark{4},
Bhupendra Acharya\IEEEauthorrefmark{5}
}
\IEEEauthorblockA{
\IEEEauthorrefmark{1}
University of Augsburg
}
\IEEEauthorblockA{
\IEEEauthorrefmark{2}
Saarland University
}
\IEEEauthorblockA{
\IEEEauthorrefmark{3}
Friedrich-Alexander-Universität Erlangen-Nürnberg
}
\IEEEauthorblockA{
\IEEEauthorrefmark{4}
Max Planck Institute for Security and Privacy
}
\IEEEauthorblockA{
\IEEEauthorrefmark{5}
University of Louisiana at Lafayette
}
}

\maketitle

\thispagestyle{plain}
\pagestyle{plain}

\begin{abstract}
The ubiquity of the video game industry and its large user base have transformed video games into complex social and economic ecosystems.
Unfortunately, this growing popularity also attracts cybercriminals who deliberately exploit game-specific mechanisms to target players. 
Despite this growing threat, cybercrime in the gaming ecosystem has received little systematic attention in prior research.

In this work, we present an empirical study of cybercrime affecting video game players, combining qualitative and observational analyses to characterize gaming-related attacks, identify common attack vectors and motivations, and examine player responses. 
Our study is based on an online survey with 57 international participants, semi-structured interviews with two confirmed victims of gaming-related cybercrime, and an analysis of 2,574 publicly available posts reporting cybercrime incidents across multiple online gaming platforms.
Our findings indicate that the theft of digital items is a prevalent motivation for attacks. We further observe that gaming-related features and services, such as item trading, team voting, and tournaments, create incentives for players to engage in risky interactions. In addition, our results highlight the targeted exploitation of weaknesses in customer support processes and reveal that certain security mechanisms provide only a false sense of protection.

\end{abstract}

\begin{IEEEkeywords}
video gaming, cybercrime, gaming abuse reports.
\end{IEEEkeywords}

\section{Introduction}
\label{sec:intro}

The global number of video game players has been steadily increasing for years \citep{Statista-NumberPCGamer-2022}, with an additional boost observed during the COVID-19 pandemic \citep{Chen-InvestigatingSecurity-2016}.
By 2025, the gaming industry had generated over 200 billion USD in annual revenue, surpassing video and music streaming and nearly doubling the combined revenues of the film and music industries \cite{henkel_gamings_2025, thompson_gaming_2023}.
Platforms such as Steam \cite{Valve-Steam-2025} recorded up to \textasciitilde 39 million concurrent active users in 2026, illustrating the vast number of individuals engaging with games on a regular basis \cite{Valve-SteamStats-2025}.

This rapid growth, while economically promising, can also lead to security oversights, as the focus on scalability and profit may outpace the implementation of robust protective measures.
Furthermore, gamers usually do not prioritize security, as it is not their primary concern during playing, which makes them more susceptible to attacks.
Common security risks include downloading and installing files from untrusted sources, often for the purpose of modifying games, adding functionality to game launchers, changing the in-game overlay or cheating, which are prevalent within the gaming community \cite{Kaspersky-ThreatReport-2023, 10.1145/3658644.3690190}.
Furthermore, gaming accounts often hold significant value, not only due to in-game items and virtual currencies, but also because they frequently contain personal data such as postal or email addresses, real names, chat histories, and even linked payment information like PayPal credentials.
The social aspects of gaming, such as in-game chat or public tournaments, can serve as attack vectors for social engineering with the purpose of manipulating players into revealing sensitive information or performing harmful actions.
Moreover, gamers often spend substantial amounts of time online, which increases their exposure to potential security threats.

In recent years, a number of high‑profile cybercrime incidents related to video games have drawn widespread media attention, underscoring how gaming communities and platforms have become significant targets for online attackers.
For example, in 2021, a vulnerability in the Java logging library Log4j was discovered, which was notably exploited in attacks against \emph{Minecraft} \cite{minecraftlog4j}.
Similarly, in 2021, criminals earned over 2 million USD using a hidden cryptominer called \emph{Crackonosh}, embedded in the code of popular games distributed for free through forums, which secretly generated cryptocurrency once the game was downloaded \cite{tidy_crackonosh_2021}.
Recently, in 2025, a demo version of the game \emph{Sniper: Phantom's Resolution} was offered on the Steam store, which was quickly identified as malware \cite{snipermalware}. Although video gaming attacks are known, there
is no systematic study that comprehensively analyzes their
operation, scale of abuse, and victims' impact.

Our research work investigates cybercrime in the video game ecosystem by identifying prevalent attack types and the domain-specific features that enable them. More specifically, we conducted a survey of 57 video game players and interviews with two confirmed victims. Building upon this qualitative work, we further collected over 16,000 posts from multiple gaming platforms, identifying more than 1,000 victims who were subjected to various types of cybercrime attacks. We further examined how players experience and respond to these incidents, including their material and emotional impacts. 

\smallskip \noindent
\textbf{Contributions.} We summarize key contributions as follows:
\begin{compactitem}
    \item We perform the most extensive study of cybercrime in the video gaming ecosystem so far, combining a mixed-method online survey, semi-structured interviews, and an analysis of user reports on cybercrime experiences.

    \item Our empirical and qualitative analysis uncovered over 13 incidents, highlighting the scale of abuse and how video gaming platforms are often overlooked.
   
    \item We contribute a labeled dataset of gaming-related cybercrime incidents that can serve as a foundation for training machine learning models and enabling large-scale automated analysis in future research.

    \item Finally, our study establishes a knowledge base that can be used to develop targeted strategies for protecting gamers and mitigating both existing and emerging video gaming–based threats.
\end{compactitem}

\section{Background}
\label{sec:back}

This section introduces the technical and structural background of the video game ecosystems.
Modern video games have evolved into highly interconnected, platform-based environments that integrate social interaction, user accounts, virtual economies, and external services.
As a result, gameplay is no longer confined to a single application but embedded in a broader digital infrastructure that manages identity, communication, monetization, and access control.

These characteristics influence player behavior and the security landscape, making it essential to understand these characteristics to contextualize the incidents discussed in this work.
The following subsections provide an overview of these core components.

\BfPara{Digital Distribution Platforms}
Digital distribution platforms (DDPs) such as Steam function as centralized hubs for purchasing, downloading, and managing video games, while also offering social features, virtual item trading, and support for user-generated content such as modifications.
In addition to Steam, several other prominent platforms exist for PC gaming, including GOG \cite{gog} and the Epic Games Store \cite{epic_2026}, as well as platform-specific stores for other devices, such as the PlayStation Store \cite{playstation}, Google Play Store \cite{Google_Play}, and Nintendo eShop \cite{nintendo}.
Beyond content distribution, DDPs operate as service platforms that mediate interactions between players, games, and developers.
Furthermore, they provide customer support channels that users can access when technical issues or security incidents arise. 

\BfPara{Virtual Items}
Virtual items are digital assets within video games, including cosmetic customizations like skins, in-game currency, characters, or equipment.
They can hold high social or economic value, especially in games with trading, marketplaces, or competitive play, making them lucrative targets for cybercriminals.
Depending on their rarity and uniqueness, some virtual items can be worth millions of USD; for example, the most expensive Counter-Strike 2 skin is currently valued at 2.5 million USD \cite{skinflow}.

\BfPara{Game Modifications}
Video game modifications (mods) are community-created changes to commercial games, usually available for free.
They range from simple interface tweaks and bug fixes to entirely new content like storylines, characters, maps and mechanics.
Mods are typically distributed directly through the DDP or third-party websites like Nexus Mods \cite{Blacktree-Nexusmods_2025}. 
Installing mods often involves running scripts or executables, sometimes requiring administrator privileges, which can be exploited by cybercriminals.

\BfPara{Cheating}
Cheating in video games refers to the use of unauthorized methods or tools to gain unfair advantages, such as modifying game files, exploiting bugs, or using external software like aimbots or wallhacks.
Such tools often manipulate game processes or memory, employing techniques similar to those used by malware.
Similar to using mods applying cheats requires users to trust third-parties.

\BfPara{Social Aspects of Gaming}
Beyond playing, gaming involves a wide range of social interactions and community activities.
These include in-game chat and voice for communication and socializing, friends lists and direct invites that maintain social connections, and community hubs and forums provided by the DDPs where players interact and share experiences.
Promotions and giveaways provide additional social engagement by attracting users, while organized tournaments, streaming, and e-sports events foster competition, collaboration, and visibility within the community.

\BfPara{Security Measures}
To protect accounts and maintain a safe gaming environment, platforms provide a range of gaming-specific security measures.
Features such as Steam Guard, platform login notifications, and account locks help monitor and prevent unauthorized access. 
Integrity checks on game files detect modified or trojanized content, while users can verify suspicious files through services like VirusTotal. 
Community-focused measures, such as chat and friend request restrictions, reporting and blocking suspicious accounts, and participating only in official tournaments or giveaways, further reduce the risk of social-engineering attacks and fraud.
\section{Related Work}
To the best of our knowledge, we present the most extensive user study of cybercrime in the video gaming ecosystem so far.
Given the extensive research on cybercrime and cybersecurity in general, in this section, we focus on previous studies specifically on cybercrime and cybersecurity in the video gaming ecosystem and highlight the novelty of our approach.

\BfPara{User Studies}
Several prior works~\cite{Feng23,Pfeiffer-GameCrime-2021,chalak2020cyber} have explored cybersecurity and cybercrime experiences in online gaming from a user perspective. Feng et al.~\cite{Feng23} performed a user study on online gaming, focusing on users from China circumventing Chinese addiction prevention systems. Pfeiffer et al. \cite{Pfeiffer-GameCrime-2021} investigate players’ perspectives on cybercrime in digital games through a focus group study.
They identify a range of offenses, including identity theft, account hijacking, credit card fraud, theft of virtual goods, and the misuse of in-game communication for harassment and other illicit activities. Similarly, Chalak \cite{chalak2020cyber} examines the prevalence of cybercrime experiences among children in online games using a survey-based approach. However, prior user studies have not comprehensively explored cybercrime across the video gaming ecosystem.

\BfPara{Developer Studies}
Klostermeyer et al. \cite{10.1145/3658644.3690190} investigate security practices and challenges in modern game development based on interviews with industry professionals. Their findings show that, although security risks are generally recognized, security measures are often deprioritized in favor of gameplay features due to time pressure, limited resources, and short development cycles. However, this work focuses on developers rather than players affected by cybercrime.

\BfPara{Cybersecurity Risks}
Several prior works have investigated cybersecurity threats, malicious activities, and structural risks within online gaming ecosystems~\cite{ibrahim_guarding_2024,chattopadhyay_role_2023,cooke_money_2024,mohr_it_2011,HIGGS2025104528}.
Ibrahim \cite{ibrahim_guarding_2024} outlines the evolving cybersecurity challenges in video games, noting that increasing online connectivity introduces threats such as data breaches, cheating, and intellectual property theft.
Chattopadhyay et al. \cite{chattopadhyay_role_2023} examine cybersecurity risks in online gaming, including data leaks and attacks causing financial losses. 
Similarly, Cooke and Marshall \cite{cooke_money_2024} investigate how video games enable money laundering, focusing on marketplaces such as the Steam Marketplace.
Mohr and Rahman \cite{mohr_it_2011} emphasize that cybersecurity is a critical concern for all organizations in the video game industry, regardless of size.
They conducted a security assessment of a leading game company to identify common risks and develop mitigation strategies.
However, this prior work has primarily focused on specific threats and technical risks rather than cybercrime in the broader video gaming ecosystem.

\BfPara{Cheating and Malware} The growing prevalence of cheats, unofficial modifications, and malicious software has motivated research into security threats within gaming platforms\cite{pontiroli2019cake,phaenthong_analyzing_2024}.
Pontiroli \cite{pontiroli2019cake} examines cheating in video games and the emerging market for cheats and cheat engines and emphasize how players may willingly execute potentially harmful software to gain in-game advantages.
Similarly, Phaenthong and Ngamsuriyaroj \cite{phaenthong_analyzing_2024} analyze security and privacy risks in popular Android games, focusing on malware, permission requests, third-party tracking, and insecure connections.
However, prior work has primarily focused on technical threats such as cheats and malware rather than broader cybercrime experiences in gaming ecosystems.

\BfPara{Novelty}
Existing research on cybercrime in gaming has examined players in small or specific subgroups, focused on developers and industry practices, or addressed isolated issues such as cheating, malware, or the misuse of games for particular crimes. In contrast, our study presents an extensive, mixed-method, player-centered analysis that combines surveys, interviews, and a systematic examination of publicly reported incidents. We explore and conceptualize cybercrime in the video gaming ecosystem, identify domain-specific attack types and vectors, analyze both material and emotional impacts on players, and uncover structural weaknesses in platforms and support mechanisms. This integrated perspective distinguishes our work from prior studies.
\section{Evaluation Setup and Methodology}
\label{sec:methodology}

\begin{figure}[th]
\centering
    \scriptsize
    \begin{tikzpicture}[node distance=2cm,on grid,auto,>=stealth]
        \node[rectangle,draw,rounded corners=7,text width=50,align=center,minimum height=25](kw){Keywords};
        \node(kw-t)[above=0.65 of kw]{User Study};
        
        \node[rectangle,draw,rounded corners=7,text width=50,align=center,minimum height=25](box1-1)[below=1 .5 of kw]{Online Survey};
        \node[rectangle,draw,rounded corners=7,text width=50,align=center,minimum height=25](box1-2)[below=1 of box1-1]{Interviews};
        \node[rectangle,draw,rounded corners=3,fit={(box1-1) (box1-2)}](box1){};
    
        \node[rectangle,draw,rounded corners=7,text width=50,align=center,minimum height=25](box2-1)[below right=0.5 and 3 of kw]{Steam\\Community};
        \node[rectangle,draw,rounded corners=7,text width=50,align=center,minimum height=25](box2-2)[below=1 of box2-1]{GOG Forum};
        \node[rectangle,draw,rounded corners=7,text width=50,align=center,minimum height=25](box2-3)[below=1 of box2-2]{Reddit};
        \node[rectangle,draw,rounded corners=3,fit={(box2-1) (box2-2) (box2-3)}](box2){};
        \node[text width=50,align=center](box2-t)[above=0.9 of box2-1]{User Report\\Collection};
    
        \node[rectangle,draw,rounded corners=7,text width=50,align=center,minimum height=25](llm)[right=3 of box2]{LLM Prompt};
        \node[text width=50,align=center](kw-t)[above=0.85 of llm]{Automated Filtration};
    
        \node[rectangle,draw,dashed,rounded corners=3,text width=75,align=center,minimum height=25](hi)[below=2.5 of box2]{Human Inspection};
    
        \node[rectangle,draw,rounded corners=3,text width=75,align=center,minimum height=25](ct)[below=1.5 of hi]{Cybercrime Tracking\\ \& Analysis};
    
        \path[->,draw] (kw) -- (1.5,0) -- node [left,circle,minimum size=10,inner sep=0,outer sep=3,draw] {2} (1.5,-1.5) -- (box2);
        \path[<-,draw] (kw) -- node [left,circle,minimum size=10,inner sep=0,outer sep=3,draw] {1} (box1);
        \path[->,draw] (box2) -- node [above,circle,minimum size=10,inner sep=0,outer sep=3,draw] {3} (llm);
        \path[->,draw] (hi) -- node [left,circle,minimum size=10,inner sep=0,outer sep=3,draw] {4} (ct);
    
        \path[->,draw] (box1) -- (0,-4) -- (hi);
        \path[->,draw] (llm) -- (6,-4) -- (hi);
    \end{tikzpicture}
\vspace*{0.1cm}
\caption{System design of our study.
}
\label{fig:sys_design}
\end{figure}
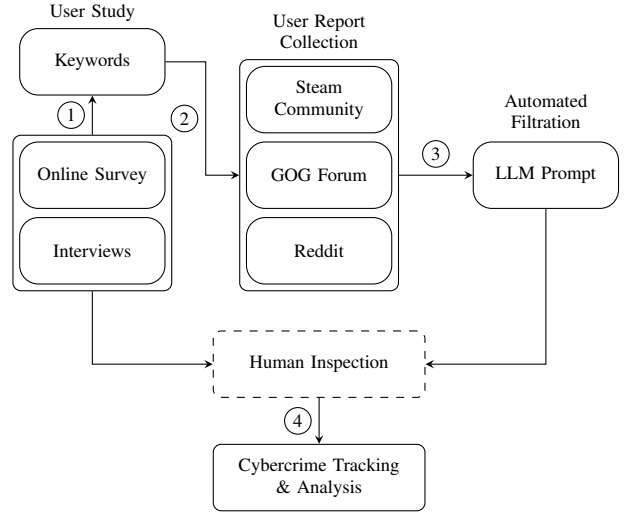

In this section, we provide detailed information on the evaluation setup, user studies, automated data collection, filtering, and human analyst review of our empirical sources. As illustrated in our system design (\Cref{fig:sys_design}), our evaluation framework consists of four main stages. First, through two user studies (i) a mixed-methods survey, and (ii) qualitative interviews, we conceptualize cybercrime in the video gaming ecosystem and derive search keywords that capture gamers’ experiences of cybercrime (\circleone). Second, we build an automated pipeline to collect incident reports from three gaming-related communication platforms based on these keywords (\circletwo).
Third, we apply an automated filtering process using LLM prompts to filter the collected data (\circlethree).
Finally, we conduct human analysis on the filtered data to perform cybercrime tracking and in-depth analysis of cybercrime incidents in the gaming ecosystem (\circlefour).
Below, we provide additional details for each stage.

\subsection{User-Study: Survey and Interviews}
\BfPara{Our motive}
We conducted a user survey to gather initial experience reports and to use the findings to identify characteristics for data scraping.
Our aim was to reach and recruit volunteer gamers for our study in order to collect information about cybercrime incidents, gaming behavior, and the communication of security risks.

\begin{figure}
    \centering
    \scriptsize
    \begin{tikzpicture}[node distance=2cm,on grid,auto,>=stealth]
        \node[rectangle,draw,rounded corners=3,align=center,text width=230,minimum height=20](quest){Survey};
    
        \node[rectangle,draw,rounded corners=3,align=center,text width=70,minimum height=20](vgu)[below left=1 and 2.8 of quest]{Video Game Usage};
        \node[rectangle,draw,rounded corners=3,align=center,text width=70,minimum height=20](ce)[below=1 of quest]{Cybercrime Experience};
        \node[rectangle,draw,rounded corners=3,align=center,text width=70,minimum height=20](demo)[below right=1 and 2.8 of quest]{Demographics};
    
        \path[->,draw] (-2.8,-0.35) -- (vgu);
        \path[->,draw] (quest) -- (ce);
        \path[->,draw] (2.8,-0.35) -- (demo);
    
        \node[rectangle,draw,rounded corners=3,align=center,text width=61,minimum height=28](ggb)[below right=1 and 0.16 of vgu]{General Gaming\\ Behavior};
        \node[rectangle,draw,rounded corners=3,align=center,text width=61,minimum height=28](mps)[below=1.1 of ggb]{Most used Platforms\\ \& Services};
        \node[rectangle,draw,rounded corners=3,align=center,text width=61,minimum height=28](uhvgm)[below=1.1 of mps]{Usge \& Handling of Video Game Modifications};
        \node[rectangle,draw,rounded corners=3,align=center,text width=61,minimum height=28](ga)[below=1.1 of uhvgm]{Gaming-related Activities};
    
        \path[->,draw] (-4,-1.35) -- (-4,-5.3) -- (ga);
        \path[->,draw] (-4,-4.2) -- (uhvgm);
        \path[->,draw] (-4,-3.1) -- (mps);
        \path[->,draw] (-4,-2) -- (ggb);
    
        \node[rectangle,draw,rounded corners=3,align=center,text width=61,minimum height=28](gsb)[below right=1 and 0.16 of ce]{General Security\\ Behavior \& \\Security Measures};
        \node[rectangle,draw,rounded corners=3,align=center,text width=61,minimum height=28](efc)[below=1.1 of gsb]{Experienced Forms of Cybercrime};
        \node[rectangle,draw,rounded corners=3,align=center,text width=61,minimum height=28](ri)[below=1.1 of efc]{Reflections of Incidents};
        \node[rectangle,draw,rounded corners=3,align=center,text width=61,minimum height=28](rsc)[below=1.1 of ri]{Received\\Security-related\\Communication};
    
        \path[->,draw] (-1.2,-1.35) -- (-1.2,-5.3) -- (rsc);
        \path[->,draw] (-1.2,-4.2) -- (ri);
        \path[->,draw] (-1.2,-3.1) -- (efc);
        \path[->,draw] (-1.2,-2) -- (gsb);
    
        \node[rectangle,draw,rounded corners=3,align=center,text width=61,minimum height=28](age)[below right=1 and 0.16 of demo]{Age};
        \node[rectangle,draw,rounded corners=3,align=center,text width=61,minimum height=28](gender)[below=1.1 of age]{Gender};
        \node[rectangle,draw,rounded corners=3,align=center,text width=61,minimum height=28](geo)[below=1.1 of gender]{Geolocation};
        \node[rectangle,draw,rounded corners=3,align=center,text width=61,minimum height=28](loe)[below=1.1 of geo]{Level of Education};
    
        \path[->,draw] (1.6,-1.35) -- (1.6,-5.3) -- (loe);
        \path[->,draw] (1.6,-4.2) -- (geo);
        \path[->,draw] (1.6,-3.1) -- (gender);
        \path[->,draw] (1.6,-2) -- (age);
        
    \end{tikzpicture}
    \vspace*{0.1cm}
    \caption{Topics of questions in the published Qualtrics survey}
    \label{fig:enter-label}
    \label{fig:questionnaires}
\end{figure}
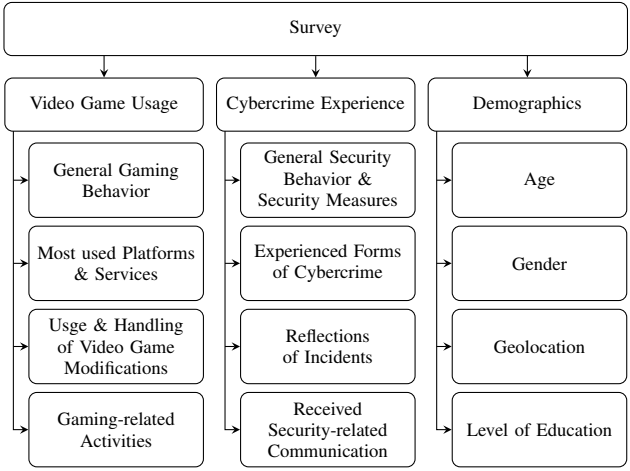

\BfPara{Survey Models}
The survey comprises 31 mixed-method questions.
The survey was designed in collaboration with experts in empirical research, taking into account accepted standards for surveys.
As illustrated in~\Cref{fig:questionnaires}, the questions are divided into three main subject areas: \emph{Video Game Usage}, \emph{Cybercrime Experience}, and \emph{Demographics}.
We provide a detailed list of the individual questions in~\Cref{sec:app:questions} and their objective in the context of this study, as follows:
\begin{itemize}
    \item \emph{\textbf{Video Game Usage}} deals with the general gaming habits and platforms used by the participants. Participants were also asked about their familiarity with third-party game modding and related security measures.
    \item \emph{\textbf{Cybercrime Experience}} focuses on participants’ general cybersecurity behavior and their experiences with cybercrime in gaming, asking them to describe incidents, resulting damage, awareness of the attack, and any subsequent behavioral changes, as well as their exposure to third-party information on cybercrime risks.
    \item \emph{\textbf{Demographics}} asks participants about their age, gender, country, and level of education.
\end{itemize}

\BfPara{Survey Distribution}
We distributed the user study survey via an anonymous Qualtrics \cite{qualtrics_nodate} link across three main channels: (i) online gaming forums based on \emph{Kaspersky’s Gaming-Related Threat Report 2023}~\cite{Kaspersky-ThreatReport-2023}; (ii) platforms associated with game distribution and modding ecosystems (e.g., Steam, GOG, and Nexusmods); and (iii) direct outreach through university mailing lists and Discord gaming communities.
These channels were chosen to maximize reach and diversity within the gaming population while targeting users likely to have relevant experiences with gaming-related cybercrime.
Before posting, we obtained permission from the official support teams or moderators of each forum and Discord server.
In total, 57 participants completed the online survey, which was distributed across 36 platforms.
We provide details of these platforms in the~\Cref{sec:app:platforms}.
All participants had to confirm that they are at least 18 years old, that they have read or have been read the declaration of consent and information on data protection that we provided, that their questions on the consent have been answered and that they are voluntarily participating in the survey.

\BfPara{Interview Models}
After completing and submitting the survey, a follow-up message was shown asking participants for their interest in an additional online video interview.
Interested participants were asked to click on a link to a second survey that only contains a single text form, asking participants to enter an e-mail address for contact, regarding the follow-up interviews.
The question was deliberately not included in the main survey 
such as access to or deletion of the address data, can be asserted individually without including the responses in the main survey. We performed semi-structured qualitative interviews with two confirmed victims of cybercrime in the video gaming ecosystem.

\subsection{User Reports: Automated Data Collection and Filtration}
\BfPara{Keywords Selection} We used the results of the user study to generate a curated set of 20 keywords spanning four thematic areas. More especially, we selected these keywords manually based on users' experienced attacks and responses. These include (i) general cybersecurity terminology (e.g., \textit{vulnerability}, \textit{exploit}, \textit{malware}, \textit{attack}, \textit{virus}); (ii) malicious behavior and bad actors e.g., \textit{hacker}, \textit{phishing}, \textit{scam}, \textit{malicious}, \textit{DDOS}); (iii) user account and identity compromises (e.g., \textit{account stolen}, \textit{account hijacked}, \textit{identity theft})  and (iv) user impacts or concerns (e.g., \textit{items stolen}, \textit{steal credit card}, \textit{experience}, \textit{security}, \textit{antivirus}).
We followed an inductive approach by first conducting a user study and then using the resulting keywords to guide quantitative automated data collection, ensuring comprehensive coverage of both attacker methods and user experiences.
The keywords used for automated data collection are listed in~\Cref{sec:app:scraping_keywords}.

\subsubsection{Automated Dataset Collection}
\BfPara{Sources} For user reports, we selected three popular gaming-related communication platforms based the number of users~\cite{ofpc_gaming_communities_2024, gog_check_these_facts_2021, Valve-SteamStats-2025}: (i) \emph{Steam Community}~\cite{SteamCommunityDiscussions}, (ii) \emph{GOG Forum}~\cite{GOGForum}, and (iii) several video game–related \emph{Reddit subforums}~\cite{Reddit}.
Subreddits were chosen for scraping based on the subreddits name and popularity. To search for appropriate subreddits we used generic gaming related terms (e.g. gaming), popular gaming devices and consoles (e.g. playstation, xbox, …) and the most played games on Steam (e.g. CS:GO). To maximize the number of reports on cybercrime we additionally used the game names from \emph{Kaspersky’s Gaming-Related Threat Report 2023}~\cite{Kaspersky-ThreatReport-2023} as subreddit search terms.
For the Steam Community and the GOG Forum, we developed an in-house Python scraper to collect posts along with user metadata, whereas for Reddit, we used the official API for content search and retrieval.
We used a substring-based keyword matching approach for scraping, allowing keywords to also match extended word forms.
Typographical errors were not explicitly addressed.
We provide the complete list of dataset subforums in~\Cref{sec:app:scraping_urls}. 

\BfPara{Raw Dataset} In~\Cref{tab:forum_dist}, we provide the breakdown of our collected dataset, which shows that 65.3\% (11,053/16,928) of the posts constitute Reddit subforums and the remaining 25.3\% (4,280/16,928) and 9.4\% (1,595/16,928) from GoG Forum and Steam Community respectively. In total, we collected 16,928 distinct posts from 12,340 distinct users.
Note that the raw dataset was pre-filtered using keyword-based scraping.

\begin{table}[h]
    \caption{Overview of the raw dataset of user reports collected from three gaming platforms.}
    \label{tab:forum_dist}
    \centering
    \begin{tabular}{l r r}
        \toprule
        \textbf{Forum} & \textbf{Posts} & \textbf{Accounts} \\
        \midrule
        Steam Community & 1,595 & 1,476 \\
        GOG Forum & 4,280 & 1,829 \\
        Reddit Subforums& 11,053 & 9,035 \\
        \midrule
        \textbf{Total} & \textbf{16,928} & \textbf{12,340} \\
        
        \bottomrule
    \end{tabular}
\end{table}

\subsubsection{Data Filtration}
\BfPara{Prompt-based Filtration} We performed LLM-based data filtration to isolate related reports of cybercrime within the video gaming ecosystem. More especially, we applied OpenAI's GPT-4o-mini model with a specialized system prompt to assign each post a binary label e.g. cybercrime vs. non-cybercrime related to the gaming environment. To mitigate LLM limitations related to task-overloading within a single prompt, we implemented three distinct iterations of filtering passes. The first pass checks whether content is relevant to both cybercrime and gaming by providing clear examples of video gaming-related cybercrimes. The next two passes each focused sequentially to check each criteria separatly to cybercrime and gaming context. This layered approach was intentionally designed to remove noise in the raw dataset and only create high-quality data for evaluation. We provide prompt engineering details in~\Cref{sec:app:scraping_promt}.

\BfPara{Manual Evaluation and Efficacy} To assess the quality of the LLM output, we manually reviewed a random sample of 200 posts across platforms and keywords. We identified that the labeling was consistent, thus reflecting a design choice that identified the relevant data.

\BfPara{Data Filtration Summary} Starting with 16,928 forum posts collected from 12,340 distinct user reports, we identified 2,574 cybercrime reports from 2,464 users, representing 15.21\% of the overall dataset from Steam Community, GOG Forums, and Reddit subforums. For the remaining quantitative analysis, we limit our findings to these 2,574 posts.

\subsection{Human Inspection and Labeling}
Following completion of the user study and automated data filtering, we conducted a qualitative analysis of the collected data to assign labels and identify cybercrime characteristics. Below, we provide further details on the manual analysis and labeling for each dataset.

\subsubsection{Labeling and Inspection of User Study}
All quantitative analyses were conducted using standard statistical methods.
The interpretation and consolidation of the rich non-numerical data material from the survey and the interviews was done using Mayring's qualitative content analysis \cite{Mayring2019}.
For this purpose, inductive categories were formed based on the research questions and responses.
Information provided in the free-text fields of the survey was analyzed individually and assigned to the relevant categories.
A list of all categories can be found in the~\Cref{sec:app:cat1}.

\BfPara{Code Formation Techniques} For each category, codes were defined based on the participants’ responses. Each code represents a distinct type of information provided. Care was taken to avoid overgeneralizing participants’ statements while maintaining a balance between capturing meaningful facets and preventing an overly detailed code set. Coding of both survey and interview responses was conducted by two of the authors.

\BfPara{Interview Transcription} The interview responses were transcribed in accordance with the transcription rules proposed by Kuckartz and Rädiker \cite{kuckartz2019transcribing}. 
The resulting statements were then analyzed using Mayring’s qualitative content analysis \cite{Mayring2019}, based on categories and codes derived from the survey responses.

\subsubsection{Labeling and Inspection of User Reports}
Following the qualitative analysis approach established in the user study phase, the interpretation of the scraped user reports was likewise conducted using Mayring’s qualitative content analysis \cite{Mayring2019}.
The categories and codes developed during the user study served as the initial analytical framework for this subsequent analysis.

\BfPara{Code Formation Techniques} Coding of the scraped data was performed by eight coders, including some authors and student helpers. All coders received detailed coding guidelines and participated in a training session conducted by a senior researcher before the coding process began. To ensure intercoder agreement, the coding approach, categories, and codes were continuously discussed and refined throughout the process as new themes emerged. Regular meetings were held to resolve ambiguities, align interpretations, and, when necessary, re-code specific posts. In addition, a senior researcher reviewed the complete coded dataset to ensure consistency and quality. Particular attention was paid to balancing sufficient detail with avoiding excessive complexity, resulting in a coding scheme suitable for exploratory research.
A list of all categories can be found in the~\Cref{sec:app:cat2}.

\BfPara{Cybercrime Tracking and Analysis} Once the user reports had been coded, standard analyses appropriate for categorical and qualitative data were performed to summarize and interpret the results.
Co-occurrences and correlations between codes were examined, and insights and patterns observed during the coding process were systematically evaluated.

More details on the individual study components, including the full list of survey questions, distribution platforms, defined categories and codes, and the keywords used for data collection, are available on GitHub: \url{https://anonymous.4open.science/r/Cybercrime-in-Video-Gaming-Ecosystem-47F6}.
\section{Delineation of User Study Responses}
Between March and May 2025, we designed and conducted an online survey, during which a total of 83 responses were collected via the Qualtrics platform. Of these, 26 responses were excluded due to incomplete answers or skipped questions. In total, 57 valid responses were retained for analysis, and we provide an analysis of these valid responses in this section. More specifically, we present (i) a statistical analysis of the survey responses, (ii) a qualitative evaluation of responses from participants who reported being victims of cybercrime, and (iii) findings from follow-up interviews conducted with two victims as follows.

\subsection{Statistics of Survey Responses}
\BfPara{Age Group} The participants’ birth years ranged from 1970 to 2004, corresponding to ages between 21 and 50 years.
21.0\% of participants did not provide their birth year.
No significant correlations were found between age and the security measures used or types of cybercrime reported.

\BfPara{Sex} 77.2\% of participants identified as male, 12.3\% as female, and 1.8\% as non-binary or third gender, while 8.8\% did not specify their gender.

\BfPara{Geographic Location} Most participants live in Germany (59.7\%).
The United Kingdom and the United States were the next most frequent (7.0\% each), followed by Brazil, Finland, Greece, Indonesia, the Netherlands, Poland, South Korea, Spain, Sweden, and Uruguay (1.8\% each).
Five participants did not provide country information (8.8\%).

\BfPara{Education} Regarding education, most participants (43.9\%) reported a high school or equivalent level, 26.3\% hold a bachelor’s degree, 17.5\% a master’s degree, and 1.8\% a doctoral degree. 8.8\% of participants did not provide educational information.

\BfPara{Platform Usage and Engagement} 64.9\% of participants stated that they regularly use chat functions to engage with others.
50.9\% use platform community features such as forums, and 21.0\% engage in official or player-organized tournaments and competitions.
15.8\% purchase in-game items from third-party platforms, and 8.8\% take part in community events.
1.8\% indicated using Discord.
17.5\% reported not engaging in any of these extended gaming activities.

\BfPara{Security Measures} Participants reported a range of security measures used in gaming and digital activities more broadly. The most common precautions included avoiding unknown attachments in emails or chats (93.0\%), using unique passwords (89.5\%) and strong passwords (86.0\%), as well as avoiding unknown hyperlinks (84.2\%). Many participants declined friend requests from unknown users (80.7\%) while still using communication and community features, suggesting that trust varies depending on the communication context. In addition, a large proportion reported technical protective measures, including two-factor authentication (2FA) (79.0\%), firewalls, trusted download sources, and regular software updates (71.9\%), as well as antivirus software (70.2\%) and avoiding risky websites (50.9\%). Less common practices included regularly changing passwords (12.3\%), while only 1.8\% reported using a hardened browser for unknown links or VirusTotal checks. Another 1.8\% reported not using any protective measures. Overall, these results suggest a high level of security awareness and the use of multiple protective strategies among participants.

\begin{figure}
    \centering
    \includegraphics[width=\linewidth]{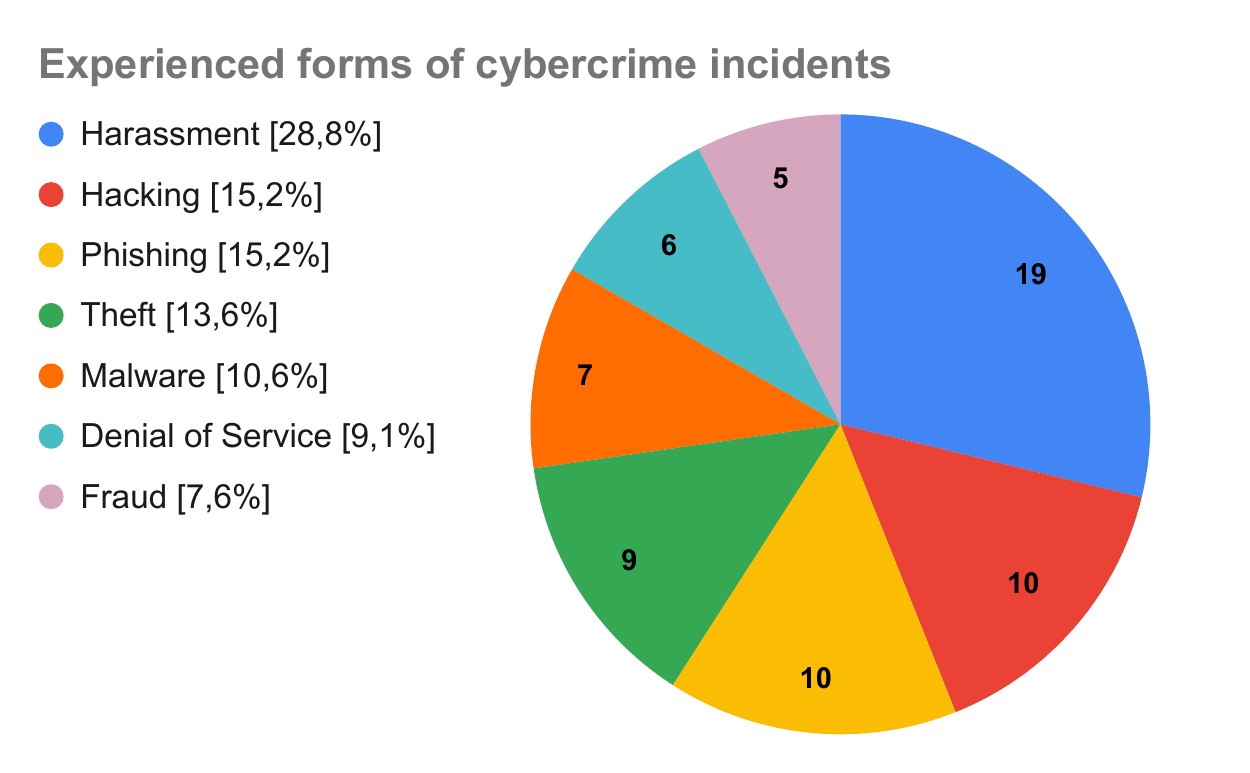}
    \vspace*{0.05cm}
    \caption{Experienced forms of cybercrime by participants (N = 34). 
    }
    \label{fig:surv:crime}
\end{figure}

\BfPara{Cybercrime Incidents and Experiences} 59.6\% of participants reported having experienced cybercrime in the context of video gaming, accounting for 66 incidents in total. Harassment was the most frequent type, representing 28.8\% of all incidents. This was followed by hacking (15.2\%), phishing (15.2\%), and theft (13.6\%).
Note that in the video gaming context the term hacking is often used as an umbrella term spanning both cybercrime-related activities and cheating, without distinguishing between contexts or methods.
Malware accounted for 10.6\% of all incidents and was reported exclusively in the context of downloading third-party game modifications. Denial-of-service attacks represented 9.1\% of incidents. Fraud was the least common category, accounting for 7.6\% of incidents. Only a small fraction of fraud-related cases involved the purchase of in-game items from third-party platforms.
A summary of the cybercrime incidents the participates experienced can be seen in~\Cref{fig:surv:crime}.

Across the 66 selected cybercrime incidents, approximately two thirds resulted in some form of loss. Account theft was the most frequent outcome (18.2\%), followed by loss of in-game currency (13.6\%), while real-money losses and loss of in-game items each accounted for 8.8\%.

Technical impacts included loss of access to files (7.0\%).  Less frequent outcomes included reduced motivation to play (5.3\%) and decreased self-confidence (3.5\%). Isolated cases included platform-related issues such as loss of balance or loss of trust in the operator (1.8\% each). 

\begin{figure}
    \centering
     \includegraphics[width=\linewidth]{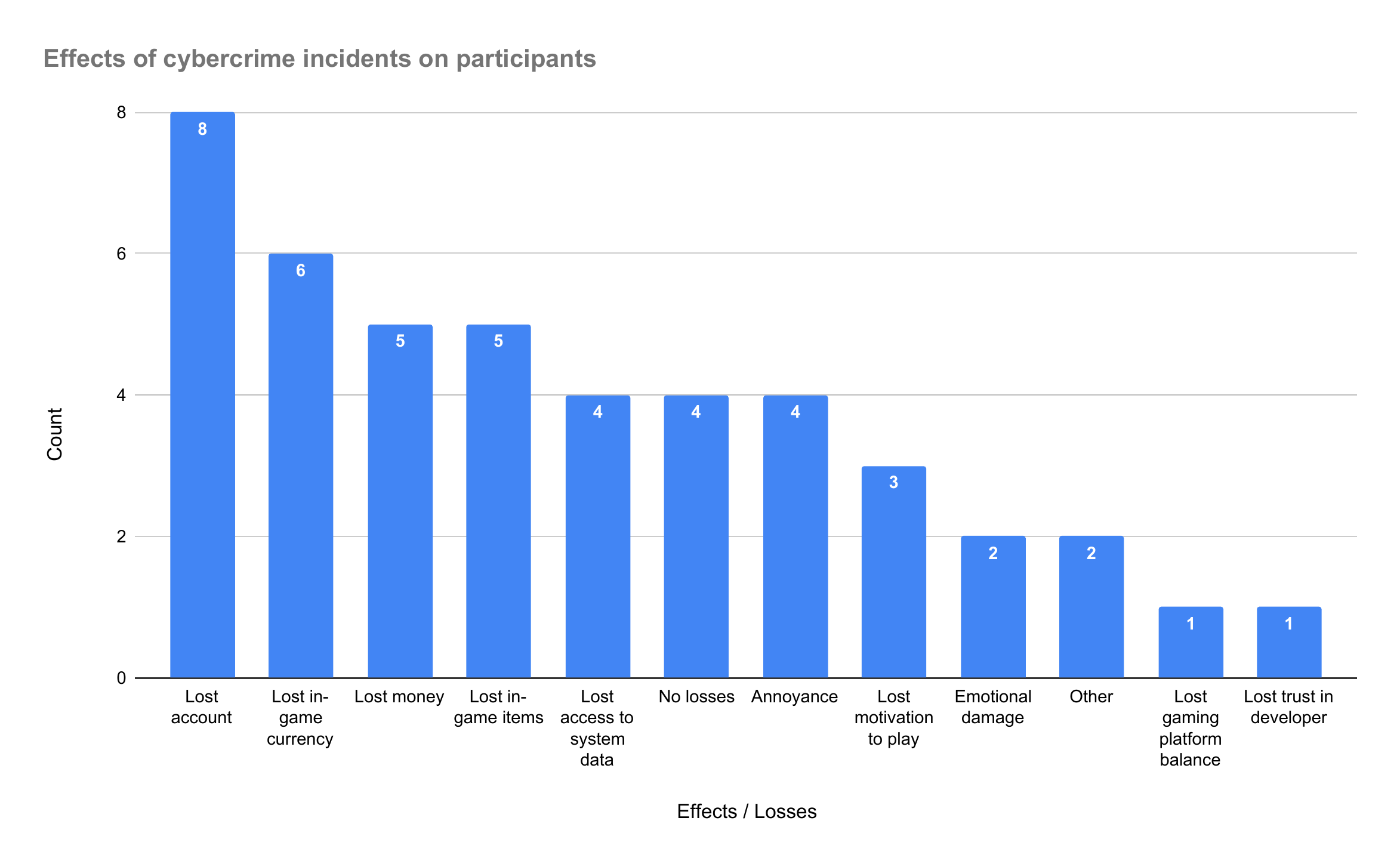}
    \caption{Effects and losses of cybercrime incidents for participants (N = 34).
    }
    \label{fig:loss}
\end{figure}

In \Cref{fig:loss} we provide the summary of the responses.

\subsection{Qualitative Analysis of Cybercrime Experiences}
52.6\% of participants provided qualitative descriptions of their cybercrime experiences, reporting a total of 81 incidents. 
Cases exhibiting characteristics of multiple cybercrime types were included in all relevant sections.
Theft was the most common incident (28.4\%), followed by Harassment (21\%) and Hacking (19.8\% cases).

Theft mainly affected PC accounts on distributor platforms, with some cases involving fraud on marketplaces.
Attack vectors included hacking, phishing, malware, weak or reused passwords, compromised e-mails, and use of other people’s devices. Fake websites, malicious links, and in-game features like items, trades, and gifting were exploited.

One participant stated:
\begin{quoting}
    \textit{``Someone (pretended) to be a Valve admin investigating hacked items and asked me to trade my skins to a safe account.''}
\end{quoting}

Detection occurred through failed logins, missing items, or platform notifications, with some cases discovered via data breaches.
Most incidents were resolved via platform support, often with account recovery. Reported impacts included financial and item loss, account compromise, and occasional loss of trust in DDPs. Participants reported improved security behavior afterward, such as stronger authentication, increased vigilance, and use of security tools.

Harassment occurred mainly in competitive PC games and on platforms like GOG forums and Twitch, often in Ranked PvP modes.
Another participant stated:
\begin{quoting}
    \textit{``During some matches of all those games I was victim of verbal harassment due to my gender. I was blamed for bad games, yelled at and misogynistic comments were made.''}
\end{quoting}
Platform features such as streaming, forums, and crowded player hubs facilitated harassment.
Effects were mostly emotional, including stress, loss of self-confidence, and reduced enjoyment. Many adjusted their gaming behavior, avoiding voice chats, playing less in competitive modes, stopping certain games, or limiting engagement in toxic communities.

Hacking is mainly used for stealing accounts or assets, usually targeting platform accounts rather than specific games. 
Attack vectors included weak or reused passwords, data breaches, and compromised emails, with no social engineering reported. Incidents were typically detected via account anomalies or platform notifications and resolved through support, often restoring access. Reported impacts included financial and item loss as well as occasional account compromise, with no emotional harm reported.
Participants improved security practices afterward, mainly through stronger authentication and increased vigilance. Hacking is perceived as a serious risk, with reused passwords and compromised devices as key vulnerabilities.

\subsection{Semi-Structure Interview with Victims}
Eight respondents provided an e-mail address and were contacted for a follow-up interview; two agreed to participate.

\BfPara{Interview 1} The participant described coordinated hacking and theft across Steam, Discord, and Humblebundle \cite{Humblebundle} accounts. 
Attackers exploited weak or reused passwords, possibly from data breaches, and bypassed 2FA.
The participant detected the incidents through unusual logins, altered account details, and missing items, responding by changing passwords, securing remaining assets, and contacting support, which recovered most accounts and refunded some losses.

Permanent losses included Steam Points and unclaimed Humblebundle keys.
The incidents caused significant emotional stress and highlighted vulnerabilities often underestimated by players, such as reused passwords and reliance on protective features that can be bypassed.

In response, the participant adopted stronger passwords, expanded use of password managers, consistently enabled 2FA across accounts, and monitored emails and downloads more carefully.
The interview underscores the risks of multi-platform credential reuse and the emotional impact of coordinated account theft, while also showing that even when protective measures fail, informed users can mitigate further damage through proactive responses.

\BfPara{Interview 2} The participant discussed cybercrime incidents primarily in their roles as a moderator on a Discord server and co-operator of a community-driven trading website. The most notable threats were denial-of-service attacks against the website, commissioned by competitors, and phishing attacks via Discord, often exploiting trust and gaming-related incentives like gift card giveaways. The participant stated:

The participant detected attacks through website downtime, suspicious messages, and compromised accounts, responding by blocking users, implementing technical protections against future attacks, and advising affected individuals on security measures. No financial or emotional harm occurred in these cases, though initial threats caused some stress.
The interview highlights risks unique to community platforms, such as attacks motivated by competition and the spread of phishing via trusted social channels. It also underscores the importance of privacy, cautious use of accounts, and user education, as security largely depends on proactive behavior rather than enforced measures.
\section{Cybercrime at Scale: User Reports Analysis}
We collected 2,574 posts as candidate posts from automated filtration. Upon performing qualitative analysis through human inspection, we excluded 1,059 posts as irrelevant. The remaining data were analyzed using co-occurrence matrices and Pearson correlations to examine associations between cybercrime types, platforms, and victim responses through a qualitative approach. During our coding, we limited ourselves to explicitly stated information, and posts could receive multiple codes across categories. The key findings are summarized below, where reported values denote code co-occurrences followed by Pearson correlation coefficients.

\subsection{Fraud and Scam}
\BfPara{Overview} Fraud and scam were the most frequently mentioned cybercrime incident (728). Fraud and scam incidents in the dataset follow recurring and well-defined patterns and are almost exclusively mediated through platform-based chat functions (317, 0.478).
Initial contact typically occurs via the DDP and is often followed by a request to continue the conversation on third-party platforms such as Discord (117, 0.267).
This shift is deliberately used to complicate later intervention by official platform support.
In the following, we use the term scam as an umbrella term encompassing all forms of fraud and deceptive practices.

\BfPara{Mechanisms and Attack Patterns} Several distinct scam variants were identified.
One of the most prevalent forms in the gaming ecosystem is the support scam, in which attackers initiate contact while impersonating official DDP support (264, 0.458), often using fake profiles that closely resemble legitimate accounts.
They create artificial urgency by claiming, for example, that the victim’s account is under investigation, and pressure the victim into transferring valuable items to a supposedly safe account that is in fact controlled by the scammer.
In some cases, victims are convinced that an account reset is required to resolve the alleged issue, further facilitating the scam.

In another form of this scam, the scammer contacts the victim and claims to have accidentally reported the victim’s account for illegal purchases or other policy violations.
The scammer then recommends contacting ``support'', which is in fact another account controlled by the attacker.
This fake support urges the victim to disclose sensitive account information or recent purchase history in order to appeal the alleged report.
Throughout the interaction, urgency and fear are deliberately induced to manipulate the victim into compliance, often resulting in account takeover or identity theft.
For these types of scam in particular, scammers are increasingly utilizing Discord (89, 0.411).

Another variant of this scam occurs when attackers already have limited access to or partial control over the account (e.g. by exploiting vulnerabilities).
In these cases, friends may be deleted or blocked (21, 0.080), and the account is manipulated to appear banned.
Fake support then demands payment to restore access, often escalating pressure by making games appear to be gradually removed from the account while continuing to communicate with the victim.

Another frequently observed pattern is the abuse of tournaments and voting systems.
In this voting scam, victims are contacted by what appears to be a friend (55, 0.168), often a compromised account, and are asked to vote (21, 0.107) for a team in a tournament (89, 0.240).
The link provided for voting leads to a phishing website that prompts the victim to log in using their DDP credentials.
A closely related variant follows the same pattern but asks victims to vote for a team logo instead.

Another tournament-related variant involves scammers contacting players after a match and asking to be added as a friend (100, 0.251), claiming that their team urgently needs an additional player for an upcoming tournament.
Victims are approached with flattery about their performance, framed as a good fit for the scammer’s team, and enticed with the prospect of winning substantial prize money in an upcoming competition.
Victims are then directed to fake tournament websites that request login with DDP credentials, resulting in account takeovers.
Trust is gradually established through friendly interaction and reinforced by time pressure and plausible narratives designed to lower the victim’s defenses.
In addition to the means of communication and the social engineering aspect, phishing websites (305, 0.461) are the most important feature of these types of scams.

Fake giveaways (62, 0.178) constitute another scam variant, in which victims are lured by messages advertising discounts or free items and are subsequently redirected to phishing websites that compromise their account credentials.
Further reported scams include fake game test offers (11, 0.059), in which victims are invited to test unreleased games and are redirected either to phishing websites or to malicious downloads.
Gambling-related (9, 0.068) scams were likewise reported, in which victims are promised guaranteed wins on item-based gambling websites if they first stake additional items, which are ultimately lost.
Item scams were also observed, where victims unknowingly purchase fake or non-existent items (9, 0.068) through fraudulent offers.
Less common but still notable are fake group (6, 0.065) invitations that primarily serve to establish contact and prepare subsequent attacks.
Another recurring pattern involves the sale of accounts (3, 0.028) containing games and wallet balances at unusually attractive prices; these accounts are typically suspended shortly after purchase because they are stolen, funded through compromised payment information, or violate platform policies prohibiting account trading, and in some cases are additionally banned for cheating.

\BfPara{Impact and Detection} Emotional responses often include feelings of insecurity (262, 0.098) and sometimes self-related distress (77, -0.027) such as shame and self-blame. Victims usually become aware of scam incidents through anomalies they observe themselves (214, -0.035). These include unexpected system behavior, suspicious actions by conversation partners, missing items or wallet balances, or disappearing friend lists. In some cases, other players (82, 0.028) alert victims that spam messages are being sent from their account, or victims recognize similarities to incidents reported by others. Login notifications (90, -0.045) indicating access attempts from unfamiliar IP addresses or at unusual times also contribute to detection. However, fraud and scam also correlate with no actual impact (297, 0.440), indicating that many users are aware of the risks of such attacks through gaming-related mechanisms like chat functions and are able to recognize and stop the attack before any harm occurs.

\BfPara{Response and Consequences} Many victims report their scam experiences in order to warn others and inform the community about new types of scams (306, 0.244).
Furthermore, scam often leads to account theft (287, -0.161) and general theft (116, -0.065) and Steam (545, 0.244) seems to be particularly affected by scams and the resulting consequences.

\subsection{Account Theft and General Theft}
\BfPara{Overview} One of the most frequently discussed incidents was account theft (724) and general theft (281), with account theft often leading to additional general theft (154).

\BfPara{Mechanisms and Attack Patterns} Account theft was reported in connection with a variety of other attacks, including scams (287, -0.161), malware infections (22, -0.128), abuse of vulnerabilities (6, 0.012), brute-force attacks (3, 0.047) and data breaches (2, -0.026).

Once an account or its credentials are compromised, the legitimate user loses access (284, 0.451) and the attacker steals its assets.
A commonly reported consequence of account theft is the loss of in-game items (142, 0.066).
Items are typically either transferred to other accounts or sold through marketplace mechanisms.
Furthermore, attackers frequently abused stored balances or linked payment information resulting in financial losses (96, 0.079).
Victims reported unauthorized purchases (52, 0.581) of games, in-game items, gift cards, or other digital goods, which were then transferred or gifted to other accounts.
In some cases, attackers purchased items of little value from themselves at inflated prices, suggesting an attempt to indirectly transfer funds.
These observations indicate that financial exploitation often extends beyond direct account access and includes more complex mechanisms for extracting value.

In several cases, stolen accounts were actively used by attackers to play games, frequently involving cheating (25, 0.781) which often leads to the account being permantly banned (94, 0.422).

Additionally, stolen accounts were sometimes used to target further victims (26, 0.124), particularly friends of the original account holder, indicating a cascading effect where trust relationships are exploited to extend the attack.
A small number of reports described stolen accounts being used to trade items (6, 0.051) back and forth between multiple accounts, which may indicate attempts at laundering digital assets.

\BfPara{Impact and Detection} Victims typically noticed attacks through anomalies, such as unexpected account activity (471, 0.037). Across many reports, a strong emotional pattern emerged among victims of account theft.
Users frequently expressed helplessness (233, 0.019), often driven by platform policies that prohibit the restoration of items or the reversal of bans, even when malicious activity is evident.
As a result, victims commonly felt punished despite being the affected party.
While anger and frustration (175, 0.079) was occasionally expressed, helplessness was the dominant emotion.
Many users turn to community or official platform forums to seek advice or assistance and to request account reviews or restoration, emphasizing their innocence and security efforts.

\BfPara{Security Measures} 2FA (122, 0.201) often fails to prevent these attacks, particularly in cases involving scams and phishing, where attackers bypass 2FA entirely. Many users also reported that, in other types of account theft, it was unclear how the attackers managed to circumvent 2FA, leaving them vulnerable despite these additional security measures. The security measures most commonly implemented following account theft were user-initiated credential resets (124, 0.209) and device deauthorization (32, 0.111).

\subsection{DDoS}
\BfPara{Overview} Players also reported of DDoS attacks (170), typically noticing them when games suddenly became unplayable (137, 0.773), servers crashed (8, 0.110), or e-sports streams (1, 0.072) could no longer be accessed.

\BfPara{Mechanisms and Attack Patterns} In some cases, DDoS attacks are not only used to disrupt availability but also as a competitive tactic, with attackers deliberately disconnecting opponents to influence the outcome of matches.
Less frequently, these attacks target supporting infrastructure, such as Digital Rights Management (DRM) servers (1, 0.045), rather than the game itself, showing that availability attacks can affect multiple layers of the gaming ecosystem.

\BfPara{Impact and Detection} DDoS attacks are often addressed in official news posts (29, 0.362), suggesting that server operators are well aware of the issue and acknowledge their responsibility in mitigating such incidents.
This differs from other types of cybercrime, where players are often expected to take more of a proactive role in prevention.
As a result, affected players experience feelings of helplessness (49, 0.088) since the disruption is largely outside their control, even though the direct impact is usually limited to annoyance and frustration (55, 0.101).
More severe emotional effects are rare, reflecting the temporary nature of these incidents, but the lack of direct recourse highlights their vulnerability.

\subsection{Infection}
\BfPara{Overview} Infection (113) is another type of attack frequently reported by players.

\BfPara{Mechanisms and Attack Patterns} These incidents are strongly associated with infected downloads (70, 0.685), indicating that malware primarily spreads through files such as plugins for game launchers or scripts that modify game overlays.
Infections can also arise from seemingly legitimate software downloads, including (fake) game clients, messengers (1, 0.034) or games (7, 0.142).
In some cases, players encounter infections when downloading game cracks (2, 0.100) to play older titles or regain access to legitimately owned games with stolen CD keys.
Interestingly, modding (17, 0.333) was mentioned less frequently than expected, though it still shows a notable correlation.

\BfPara{Impact and Detection} Rarely, infections can escalate to turn systems into bots (5, 0.203) or render them completely unusable (2, 0.128), demonstrating that the severity of impact varies widely.
Victims typically report feelings of insecurity (24, -0.061) and irritation (20, -0.022), while detection often occurs through antivirus alerts (31, 0.483).
Following an infection, players frequently take action such as resetting system components (20, 0.291) to mitigate potential damage.

\BfPara{Response and Platform Impact} Among platforms, GOG (13, 0.189), particularly GOG Galaxy, appears more frequently affected than other digital distribution platforms.

\subsection{Harassment}
\BfPara{Overview} Harassment was mentioned in 26 posts and is considered only when players are directly targeted within the gaming environment (specifically excluding third-party communication platforms such as Reddit).

\BfPara{Mechanisms and Attack Patterns} This includes not only verbal abuse in in-game chat (11, 0.062) but also technically enabled harassment, such as taking control of a player’s character, manipulating the game world or inventory, or using another character to harass the victim in-game.
While such cases are uncommon in our dataset, they are reported as highly distressing.
Similar phenomena are well documented outside our data, particularly on platforms such as Roblox (2, 0.042), but also occur across other games. In some cases, this resulted in a reduced usage of video games (3, 0.216).

Insult-based harassment via the in-game chat is largely absent from our dataset, potentially due to missing keywords during scraping or because such behavior has become normalized and is therefore less frequently reported.

\BfPara{Impact and Detection} Psychological damage (6, 0.387) occurs exclusively in harassment-related incidents, with fear (11, 0.032) and anger (11, 0.070) being the dominant emotional responses. Harassment shows strong correlations with psychological damage (0.387, 0.470 and 0.332), highlighting the significant emotional impact of these attacks. Extreme forms such as doxing (2) and swatting (1) were rare but particularly severe, illustrating that even infrequent incidents can have a disproportionate effect on victims’ safety and well-being.

\subsection{Role of Support}
\BfPara{Overview} Our dataset shows that the majority of reported issues were either not resolved (186) or only partially resolved (45) by support, while effective assistance was provided in just 54 cases.

\BfPara{Impact and User Experience} Unresolved or partially resolved incidents frequently trigger strong emotions among affected users, including anger and frustration (96, 0.284) as well as disappointment (43, 0.136), and in some cases may even lead to boycotts of the platform (7, 0.118).

\BfPara{Limitations and Restrictions} Support difficulties are often compounded by regional restrictions, with victims being told that their accounts fall outside the service region. 
Attackers sometimes deliberately exploit these limitations by targeting accounts in specific regions or transferring accounts to areas where support is limited.

Additional complications arise from company policies and procedural constraints.
For example, support may be unable to restore stolen virtual items if doing so would compromise their value due to duplication.
Reporting account theft can be particularly challenging on platforms like Steam, as opening a support ticket typically requires account access, which victims lack after a theft.

In at least two cases, attackers impersonated victims to contact support directly and modify account details, bypassing verification procedures, which led to financial losses and heightened feelings of helplessness.

In one case, a subscription could not be canceled and was actively used by the attacker who had stolen the account while the victim waited for a response from support.

\BfPara{Impacts} As a result, players increasingly report a lack of confidence that platforms, developers, manufacturers, or support services can effectively protect them or provide meaningful assistance.
Frustration is often directed not at the attackers but at the official parties from whom players expect help, yet frequently receive insufficient, delayed, or no assistance.

These observations highlight that support on gaming platforms frequently fails to provide adequate assistance and that attackers actively exploit these shortcomings.
While all platforms would benefit from improvements, PlayStation appears particularly affected (0.126).

\subsection{Security Measures}
\BfPara{Post-incident Behavior} A significant portion of victims implement enhanced security measures following a cybercrime incident. Common actions include enabling 2FA, resetting credentials, deauthorizing devices, or running antivirus scans. However, these protective measures often prove insufficient against threats in the gaming ecosystem.

\BfPara{Limitations in Practice} Many players reporting account theft had 2FA enabled, yet this did not prevent phishing attacks or unauthorized access, highlighting the limitations of this mechanism in practice. Similarly, antivirus notifications regarding potentially malicious game files are sometimes ignored or dismissed as false positives, with some users even temporarily disabling their antivirus software. This may be partly due to the frequent occurrence of false positives in the past, which has led to a form of desensitization. At the same time, community advice often encourages players to run security programs despite warnings, demonstrating a reliance on peer guidance rather than official protective mechanisms.

Overall, while victims take security seriously after incidents, the data suggests that current protective measures provide limited real-world defense, and additional strategies or improved platform-level protections may be necessary to effectively mitigate risks.

One user reports on an incident where their account got stolen 6 times in 24 hours due to an email reset exploit, were two factor authentication was disabled by adding a new email address.
The 2FA is flagged as enabled within the account, but in reality, it is disabled.
This leads to a situation where attackers can access and change the account without triggering the 2FA notification.
Support only helped restore the account after invoices for the games and other evidence were provided to prove that the account actually belonged to the player and contained games.
However, after the account was stolen again through the email reset exploit, the player was simply ignored by support.

\subsection{Other Findings}
Other findings include that Steam was by far the most affected platform (948), reflecting its large user base and central role in the video gaming ecosystem.

Attacks predominantly target players of competitive (340), online (468), multiplayer (472) games, while casual (online multiplayer) games (54) are less frequently affected.
Single-player (51) and offline games (2) are rarely impacted, indicating that cybercrime in the gaming context is strongly linked to social interaction and competition.
This distribution suggests that attackers focus on environments where player interactions, in-game economies, or competitive outcomes provide the greatest potential for exploitation.

\begin{figure}
    \centering
    \includegraphics[width=\linewidth]{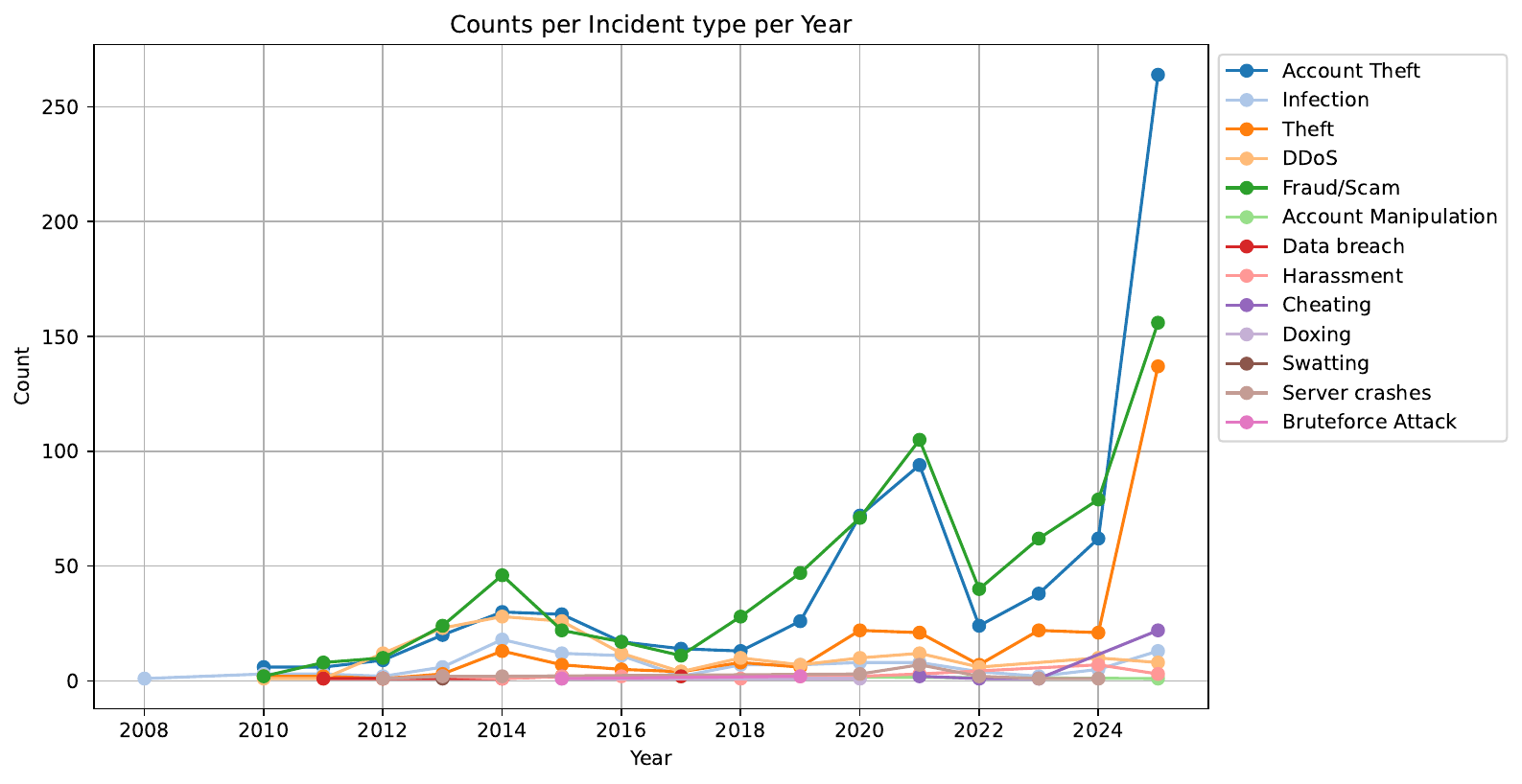}
    \caption{Distribution of reports of specific cybercrime incidents over time.}
    \label{fig:incidents}
\end{figure}

The distribution of reports of specific cybercrime incidents over time is shown in~\Cref{fig:incidents}.

\section{Discussion}

Overall, cybercrime in the video gaming ecosystem emerges as a socio-technical phenomenon driven by platform design, social dynamics, and technical vulnerabilities. 

\BfPara{Lessons Learnt} We highlight three key lessons as follows.
\begin{compactitem}
\item Cybercrime in gaming is not merely a reflection of general online crime but is driven by game-specific features, social structures, and in-game economies that attackers deliberately exploit.
\item Fraud and scams dominate the threat landscape, particularly account theft targeting valuable virtual assets, while less frequent attacks (e.g., DDoS, malware, and severe harassment) expose additional risks.
\item Attackers combine technical and social tactics to exploit gaming-specific assets, while limited platform support leaves players reacting to harm with lasting financial and emotional consequences.
\end{compactitem}

\smallskip

\BfPara{Limitations} 
Our study has several limitations. First, our dataset is subject to sampling and reporting biases, with stronger representation from PC gaming communities than console or mobile platforms. Second, many reports are subjective and one-sided, which may affect reliability and generalizability. Third, our de-duplication relied on platform identifiers and may not detect users with multiple accounts across platforms. Finally, our survey population was primarily recruited through major gaming communities, limiting demographic diversity. Nevertheless, combining surveys, interviews, and large-scale user reports enabled us to capture diverse cybercrime experiences across the video gaming ecosystem.

\smallskip

\BfPara{Future Work}
Future research can build on this work by conducting targeted studies to test specific hypotheses and leveraging the labeled dataset for automated large-scale analysis. In particular, the dataset could be used to train BERT-based models to identify and classify gaming-related cybercrime incidents at scale, enabling analyses that go beyond manual coding and support larger studies of cybercrime in gaming ecosystems. In addition, future work may refine keyword-based data collection approaches, while further surveys, interviews, and studies of underrepresented platforms (e.g., mobile gaming) are needed for a more comprehensive view of cybercrime in gaming.
\section{Conclusion}
\label{sec:conclusion}

To the best of our knowledge, this work presents the first user-centered survey study of this size on cybercrime in the video game ecosystem.
Our results indicate that cybercrime in the video gaming ecosystem is widespread and affects a substantial proportion of players.
While many affected players report employing common security measures, these mechanisms often fail to prevent attacks or mitigate their consequences.
At the same time, players demonstrate a high level of awareness regarding prevalent attack techniques.
However, attackers continuously adapt their strategies and deliberately exploit domain-specific characteristics of the gaming ecosystem to conduct targeted attacks.
These findings suggest DDPs must address cybercrime as a structural issue and strengthen technical safeguards, detection mechanisms, and support processes to provide effective protection.

\section*{Use of AI}
During the preparation of this work, AI technologies were used to assist in the writing process.
The authors used ChatGPT, Grammarly, and DeepL to revise most sections of the paper for typographical errors, grammatical correctness, and rephrasing.
After using these tools, the manuscript was carefully reviewed and the content was edited as needed.
No tools or services were used for content generation.
The authors take full responsibility for the content of this manuscript.

\section*{Ethical Considerations}
In this section, we elaborate on the Ethical Considerations of this study, as it relates to human subjects.
We consulted our institution’s Empirical Research Group to ensure the survey adhered to ethical guidelines.
Taking into consideration that cybercrime-related surveys may cause emotional distress or potential harm, the survey was jointly designed with the Empirical Research Group, psychologists, and subject-matter experts, who supported the formulation and review of the questions.
More specifically, we avoided asking sensitive questions, stored data in an authorized access protection level, and upheld integrity.
We refrain from identifying participants' information, such as names or personal references.
Participants received a clear study description and were informed they could withdraw anytime. We provided additional contact details for future inquiries.
Prior to survey conduct, we informed to participants that they can skip online Qualtrics or interview questions that they do not want to respond to, and they can stop whenever they want.
We provided follow-up contact information to video game player participants, should they have additional questions or require further clarification.
On average, participants spent 1.023,1 seconds completing the Qualtrics survey.
Before distributing the survey we also filled an Ethical Evaluation Request to the Ethical Review Board of (details are removed for the review process) for the whole research project including the survey, interview and data scraping.
The Ethical Review Board came to the conclusion that ``There are no ethical concerns against the implementation of the proposal''.

\section*{Open Science}
We support open science and accordingly, release the full artifacts of data and code, including the raw dataset collected from all three sources and other relevant data related to the study, through this anonymous GitHub link: 
\url{https://github.com/janineschneider/Cybercrime-in-Video-Gaming-Ecosystem}. 

\bibliographystyle{plain}
\bibliography{bib}

\startappendices
\clearpage

\section{Survey Questions}
\FloatBarrier
\label{sec:app:questions}

\begin{table}[h!]
\caption{Video game usage questions.}
\centering
\begin{tabularx}{\linewidth}{l|X}
   \textbf{ID} & \textbf{Question} \\
   \hline
   Q1 & How many hours per week do you spend playing video games approximately? \\
   \hline
   Q2 & Do you play as a hobby or as a profession?  \\
   \hline
   Q3 & In which year did you start playing video games?  \\
   \hline
   Q4 & What is your most used video game platform in the last 12 months? \\
   \hline
   Q5a & Which operating system do you use most to play on your PC?  \\
   \hline
   Q5b & Which operating system do you use most to play on your Smartphone/Tablet? \\
   \hline
   Q5c & Which handheld do you use most to play? \\
   \hline
   Q5d & Which operating system version do you use? \\
   \hline
   Q5e & Which PlayStation version do you use most to play? \\
   \hline
   Q5f & Which Xbox version do you use most to play? \\
   \hline
   Q5g & Which Nintendo console do you use most to play? \\
   \hline
   Q6 & Do you use a custom firmware that, for example, enables modding? \\
   \hline
   Q7 & Please select which of the following digital game distribution platforms you use the most. \\
   \hline
   Q8 & Have you ever installed third-party video game modifications? \\
   \hline
   Q9 & How do you ensure these modifications are safe to download and install? \\
   \hline
   Q10 & Please select which of the following game modification distribution platforms you use the most \\
   \hline
   Q11 & Which of the following gaming-related activities do you engage in? \\
   \hline
   Q12 & Which of the following statements describes you best? 
\end{tabularx}
\end{table}

\vspace*{0.1cm}
\begin{table}[h!]
\caption{Demographics.}
\centering
\begin{tabularx}{\linewidth}{l|X}
   \textbf{ID} & \textbf{Question}\\
   \hline
   Q22 & What is your year of birth? \\
   \hline
   Q23 & Please select the gender you identify with. \\
   \hline
   Q24 & Which country do you currently live in? \\
   \hline
   Q25 &What is your highest level of education?
\end{tabularx}
\end{table}

\begin{table}[h!]
\caption{Cybercrime experience questions.}
\centering
\begin{tabularx}{\linewidth}{l|X}
   \textbf{ID} & \textbf{Question}\\
   \hline
   Q13 & What measures do you take to protect yourself against cybercrime incidents in general?\\
   \hline
   Q14 & Have you ever been a victim in one of the following cybercrimes in the context of gaming?\\
   \hline
   Q15 & Please describe the cybercrime incidents you have encountered, including the context in which they occurred, as detailed as possible, including the game, platform, client, operating system, activity at the time of the incident, etc.. \\
   \hline
   Q16 & Please describe how you became aware of the fact that you had become a victim of cybercrime.\\
   \hline
   Q17 & What impact did these incidents have on you?  \\
   \hline
   Q18 & After these incidents, did you change your behavior in any way? \\
   \hline
   Q19 & Have you ever been informed about security risks or a security incident related to playing video games (for example, by a digital game distribution platform or a game publisher)?\\
   \hline
   Q20 & Please describe as detailed as possible how you were informed, e.g., about which incident, by whom, via which medium, the content of the notification, etc.. \\
   \hline
   Q21 & How often have you been a victim of video game-related cybercrimes? \\
\end{tabularx}
\end{table}

\newpage

\section{User Report Collection Keywords}
\FloatBarrier
\label{sec:app:scraping_keywords}

\begin{table}[h!]
\caption{Scraping Keywords}
\centering
\begin{tabularx}{\linewidth}{l|X}
\textbf{ID} & \textbf{Keywords} \\
\hline
K1 & Vulnerability \\
K2 & Exploit \\
K3 & Attack \\
K4 & Steal \\
K5 & Malicious \\
K6 & Hacker \\
K7 & Malware \\
K8 & DDoS \\
K9 & Experience \\
K10 & Security \\
K11 & Scam \\
K12 & Phishing \\
K13 & Hack \\
K14 & Account stolen \\
K15 & Account hijacked \\
K16 & Virus \\
K17 & Antivirus \\
K18 & Items stolen \\
K19 & Steal credit card \\
K20 & Identity theft \\

\end{tabularx}
\label{tab:keywords}
\end{table}

\newpage

\section{Survey Distribution}
\FloatBarrier
\label{sec:app:platforms}

\begin{table}[h]
\caption{Platforms and forums used for survey distribution.}
\centering
\begin{tabularx}{\linewidth}{X|X}
   \textbf{Platform} & \textbf{Date of Publication} \\
   \hline
   GameFAQs & 23.03.25 \\
   \hline
   Nexusmods & 23.03.25 \\
   \hline
   Reddit: r/Diablo4 & 23.03.25 \\
   \hline
   Reddit: r/Dota2 & 23.03.25 \\
   \hline
    Reddit: r/Minecraft & 23.03.25 \\
   \hline
   Reddit: r/Mobilegaming & 23.03.25 \\
   \hline
   Reddit: r/Pubg & 23.03.25 \\
   \hline
   Reddit: r/SampleSize & 23.03.25 \\
   \hline
   Reddit: r/Steamscams & 23.03.25 \\
   \hline
   Reddit: r/Survey & 23.03.25 \\
   \hline
   Steam & 23.03.25 \\
   \hline
   The Student Room & 24.03.25 \\
   \hline
   University Mailing List & 25.03.25 \\
   \hline
   GOG & 10.04.25 \\
   \hline
   Discord Server: Apex Legends & 16.04.25 \\
   \hline
   Discord Server: Hogwarts Legacy & 16.04.25 \\
   \hline
   Discord Server: League of Legends & 16.04.25 \\
   \hline
   Discord Server: PlayStation & 16.04.25 \\
   \hline
   Discord Server: Roblox & 16.04.25 \\
   \hline
   Discord Server: Sanctuary - Diablo 4 Community & 16.04.25 \\
   \hline
   Discord Server: Xbox-Now & 16.04.25 \\
   \hline
   Discord Server: World of Warcraft & 16.04.25 \\
\end{tabularx}
\end{table}

\newpage

\section{User Report Collection Subforums}
\FloatBarrier
\label{sec:app:scraping_urls}

\begin{table}[H]
\caption{Subforums used for collecting user reports.}
\centering
\begin{tabularx}{\linewidth}{X|X}
\textbf{URL} & \textbf{\# Collected Posts}\\
\hline
r/dota2 & 2261 \\
\hline
r/leagueoflegends & 1535 \\
\hline
r/gaming & 996 \\
\hline
r/steam & 994 \\
\hline
r/csgo & 792 \\
\hline
r/minecraft & 723 \\
\hline
r/apexlegends & 587 \\
\hline
r/pcgaming & 576 \\
\hline
r/diablo4 & 542 \\
\hline
r/wow & 494 \\
\hline
r/steamscams & 419 \\
\hline
r/roblox & 348 \\
\hline
r/nintendoswitch & 310 \\
\hline
r/playstation & 260 \\
\hline
r/pubg & 86 \\
\hline
r/xboxseriesx & 79 \\
\hline
r/mobilegaming & 51 \\
\hline
gog/forum/general\_archive/ & 1695 \\
\hline
gog/forum/general/ & 1541 \\
\hline
gog/forum/general\_de/ & 496 \\
\hline
gog/forum/general\_beta \_gog\_galaxy\_2.0/ & 476 \\
\hline
gog/forum/general\_pl/ & 35 \\
\hline
gog/forum/general\_ru/ & 23 \\
\hline
gog/forum/general\_fr/ & 14 \\
\hline
steam/discussions/forum/1 & 355 \\
\hline
steam/discussions/forum/9 & 346 \\
\hline
steam/discussions/forum/0 & 188 \\
\hline
steam/discussions/forum/12 & 165 \\
\hline
steam/discussions/forum/10 & 154 \\
\hline
steam/discussions/forum/7 & 94 \\
\hline
steam/discussions/forum/30 & 71 \\
\hline
steam/discussions/forum/11 & 45 \\
\hline
steam/discussions/forum/17 & 28 \\
\hline
steam/discussions/forum/24 & 28 \\
\hline
steam/discussions/forum/14 & 21 \\
\hline
steam/discussions/forum/8 & 18 \\
\hline
steam/discussions/forum/2 & 17 \\
\hline
steam/discussions/forum/13 & 16 \\
\hline
steam/discussions/forum/15 & 12 \\
\hline
steam/discussions/forum/29 & 12 \\
\hline
steam/discussions/forum/18 & 9 \\
\hline
steam/discussions/forum/16 & 7 \\
\hline
steam/discussions/forum/20 & 6 \\
\hline
steam/discussions/forum/27 & 2 \\
\hline
steam/discussions/forum/26 & 1 \\
\end{tabularx}
\end{table}

\section{User Report Filtering Prompt}
\FloatBarrier
\label{sec:app:scraping_promt}
\BfPara{LLM filtering prompt iteration 1}

You are a tool that receives an online forum post, and should decide whether it is relevant for a study on cybercrime and abuse in video games.

You will receive the data in JSON format, with attributes "id" and "text". I expect your result in the following valid JSON format without code formatting, and nothing else:
\{"id": "XYZ", "result": "ABC"\}

You can directly abort the rest of this prompt and return with "False" in the above format if the post has no connection to video games, video game platforms, online gaming communities, or gamer-related experiences.
If not, please use the following rules to determine the significance:

Return "False" if the post is only about the following OR does not deal with video game/gaming-related content at all:
- Promotional announcements, sale events, or general FAQs—even if they warn about “scams”—when no user is reporting a concrete incident of fraud or malware
- Generic Q\&A or tech-support threads without real evidence of cybercrime or abuse (e.g., “game won’t launch,” “driver issues,” or “antivirus flagged harmless game files” where no actual malware is demonstrated)
- Game-mechanics issues: bugs, crashes, performance, balancing, matchmaking, modding questions, piracy
- In-game cheating, exploits, or modding: always False unless there’s evidence of malware, phishing, or account compromise
- Discussions of fictional in-game violence, sexual content, or narrative critique that do not involve cybersecurity risks
- User complaints / excuses about being banned for cheating / hacking in-game without any suggestion of unauthorized hacking or phishing
- Any content that is completely unrelated to cyber crime / abuse

Return "True" if the post includes at least one of the following AND is clearly within a video game/gaming-related context:
- Cybercrime involving games: hacking accounts, phishing other players, exploiting game servers, in-game item scams or fraud, unauthorized third-party cheats that manipulate game code, account theft or resale, and similar malicious behavior
- Malware targeting gamers: viruses, trojans, keyloggers, spyware / adware infecting gaming PCs or consoles, infected game installers, or similar
- Suspicious software or security flags: questions about anticheat software being flagged as malware, DRM tools behaving like spyware, or security software mistakenly quarantining game clients—but only if there is clear suspicion of a genuine security risk rather than a routine support request
- Security breaches or account compromises: data leaks of gamer credentials, credential stuffing against gaming accounts, reports of forced password resets or 2FA exploits affecting players
- Unauthorized access or manipulation: unexplained in-game transactions, trades, or asset transfers that suggest someone gained illicit access to an account
- Plausible evidence of security risks for games or gaming accounts (e.g., posting virus scan logs showing game files infected with malware)
- Hate speech, harassment, or targeted abuse within a gaming community (e.g., slurs or threats directed at players in a gaming context)

If unclear, lean toward "True" only if there's a real chance of security risk (from the above "true" cases). Not just cheating or security unrelated posts.
Before you answer, make sure you fully understand what the author is reporting and whether it has a clear connection to cybercrime or malicious abuse in the video game context.\\

\BfPara{LLM filtering prompt iteration 2}
You are a tool to help a study about cybercrime / abuse within the gaming community.
You should classify if the following text obviously deals with the gaming environment.
Meaning if it is obviously about video games, the gaming scene, or gaming environment or similar.
If it is, answer with YES, if not with NO.\\

\BfPara{LLM filtering prompt iteration 3}
You are a tool to help a study about cybercrime / abuse within the gaming community.
You should classify if the following text obviously deals with actual performed cybercrime or abuse.
Meaning, if it is obvious, that the text deals about a user report of an actual performed cybercrime / abuse.
If it is, answer with YES, if not with NO.\\

\section{User Study Labeling Categories}
\FloatBarrier
\label{sec:app:cat1}

The categories used for the user report labeling are listed as follows:
\begin{compactitem}
    \item \textit{\textbf{Types of Cybercrime}} is used to classify the incidents described into the corresponding types of attacks. 
    \item \textit{\textbf{Context of Cybercrime Incidents}} is used to describe the environment in which the attacks took place. This includes the platform, the game or account, and the game mode in which the incidents occurred. If third-party modifications or websites were involved in the attacks, these are also classified. 
    \item \textit{\textbf{Attack Vectors}} provides information about the means used by the attackers in the incidents, including social engineering and/or technical measures. Gaming-related topics that were relevant to the attack are also listed. 
    \item \textit{\textbf{Incident Detection}} describes how participants became aware of the attacks, either themselves, through official notifications, or other players. If specified, the immediate response after the realization of an attack is also indicated.
    \item \textit{\textbf{Effects of Cybercrime Incidents}} describes any losses incurred by participants as a result of attacks and whether these were of a monetary, emotional, digital, account-related, and/or temporary nature.
    \item \textit{\textbf{Help and Support}} lists information provided by participants on whether official or community assistance was provided in resolving the incidents, or whether support was not sought or was not available.
    \item \textit{\textbf{Changes in Behavior after Cybercrime Incidents}} describes any adjustments or changes in the security or gaming behavior of participants as a result of the attacks.
    \item \textit{\textbf{Information about Risks}} deals with security-related communication of risks or recommendations that participants have received. These are classified according to the type of information, the source, and the medium through which the communication took place.
    \item \textit{\textbf{User Awareness of Risks}} deals with the classification of cybercrime risks and protective measures and whether these are considered dangerous, acceptable, or, in the case of protective measures, effective or ineffective. Information provided by participants who demonstrate a deeper technical understanding of the attacks, as well as underestimated risk aspects, is also taken into account.
\end{compactitem}

\section{User Report Labeling Categories}
\FloatBarrier
\label{sec:app:cat2}

The categories used for the user study labeling are listed as follows:
\begin{compactitem}
    \item \textit{\textbf{Type of post}} is used to differentiate between experience reports, questions, answers, recommendations and warnings.
    \item \textit{\textbf{Incident type}} is used to describe the incident that the player reports on.
    \item \textit{\textbf{Attack vectors/risks}} specifies the means or mechanism used to carry out the attack. These can be technical or non-technical, or relate to specific game-related elements.
    \item \textit{\textbf{Impact}} describes any impact the incident had on the player, such as monetary, digital, and account-related losses, emotional damage, leaked credentials, but also boycott or reduced usage of certain games or platforms.
    \item \textit{\textbf{Incident detection}} describes how players became aware of the incident, either themselves, through some kind of notifications, or other people.
    \item \textit{\textbf{Emotion}} describes in what state of mind the player wrote the post or how they felt when dealing with the incident.
    \item \textit{\textbf{Platform, Device, OS, Game, Game mode}} describes the context of the incident and the environment in which the incident took place.
    \item \textit{\textbf{Security measures (pre-incident)}} is used to measure which security measures (traditional and gaming related) were used as protective measures before the incident happened.
    \item \textit{\textbf{Security measures (post-incident)}} is used to measure which additional security measures players applied after the incident happened.
    \item \textit{\textbf{Support}} describes the level of help players got from the official platform support.
\end{compactitem}

\end{document}